\documentclass[10pt,journal,compsoc]{IEEEtran}
\usepackage{url}
\usepackage{cite}
\usepackage{subfigure}
\usepackage{graphicx}
\usepackage{float}
\usepackage{amsmath}
\usepackage{bm}
\usepackage{multirow}
\usepackage{multicol}
\usepackage{color}
\usepackage{array}
\usepackage{algorithm}
\usepackage{algorithmic}
\usepackage{caption}
\usepackage{amsfonts}
\usepackage[pagebackref=false,breaklinks=true,letterpaper=true,colorlinks,bookmarks=false]{hyperref}
\usepackage{color}
\usepackage{diagbox}
\usepackage{amsmath}
\usepackage{ragged2e}

\ifCLASSINFOpdf
\else
\fi
\begin{document}
%
\title{Dual Shape Guided Segmentation Network for Organs-at-Risk in Head and Neck CT Images}
%
%
%
%

\author{Shuai~Wang,
        Theodore~Yanagihara,
        Bhishamjit~Chera,
        Colette~Shen,
        Pew-Thian~Yap,
        and Jun~Lian 
\IEEEcompsocitemizethanks{\IEEEcompsocthanksitem This work was supported in part by NIH Grant CA206100 (Corresponding author: Jun Lian).

\IEEEcompsocthanksitem S. Wang and P. Yap are with the Department of Radiology and Biomedical Research Imaging Center, University of North Carolina at Chapel Hill, Chapel Hill, NC 27599 USA.
\IEEEcompsocthanksitem T. Yanagihara, B. Chera, C. Shen, and J. Lian are with the Department of Radiation Oncology, The University of North Carolina at Chapel Hill, Chapel Hill, NC 27599 USA.}
}

%
%

\markboth{}%
{Shell \MakeLowercase{\textit{et al.}}: Bare Demo of IEEEtran.cls for Computer Society Journals}
%



\IEEEtitleabstractindextext{%
\begin{abstract}
\justifying The accurate segmentation of organs-at-risk (OARs) in head and neck CT images is a critical step for radiation therapy of head and neck cancer patients. However, manual delineation for numerous OARs is time-consuming and laborious, even for expert oncologists. Moreover, manual delineation results are susceptible to high intra- and inter-variability. To this end, we propose a novel dual shape guided network (DSGnet) to automatically delineate nine important OARs in head and neck CT images. To deal with the large shape variation and unclear boundary of OARs in CT images, we represent the organ shape using an organ-specific unilateral inverse-distance map (UIDM) and guide the segmentation task from two different perspectives: direct shape guidance by following the segmentation prediction and across shape guidance by sharing the segmentation feature. In the direct shape guidance, the segmentation prediction is not only supervised by the true label mask, but also by the true UIDM, which is implemented through a simple yet effective encoder-decoder mapping from the label space to the distance space. In the across shape guidance, UIDM is used to facilitate the segmentation by optimizing the shared feature maps. For the experiments, we build a large head and neck CT dataset with a total of 699 images from different volunteers, and conduct comprehensive experiments and comparisons with other state-of-the-art methods to justify the effectiveness and efficiency of our proposed method. The overall Dice Similarity Coefficient (DSC) value of 0.842 across the nine important OARs demonstrates great potential applications in improving the delineation quality and reducing the time cost.
\end{abstract}

}

\maketitle

\IEEEdisplaynontitleabstractindextext

%
\IEEEpeerreviewmaketitle

\IEEEraisesectionheading{\section{Introduction}\label{sec:introduction}}
\IEEEPARstart{I}{n} the United States, head and neck cancers account for approximately 4\% of all cancers, and an estimated 65,630 people would develop head and neck cancers in 2019~\cite{ACS2020}. For the diagnosis and treatment of head and neck cancers, radiation therapy is an important tool and often used in combination with other therapies, such as surgery, chemotherapy, and targeted therapy~\cite{zorat2004randomized}. To optimize the dose distribution of radiation therapy, normal organs (organs-at-risk, OARs) should be accurately segmented to prevent damage. In the head and neck area, the OARs are densely distributed and some are close to the tumor to be treated, which makes them more vulnerable to irradiation. In the standard protocol, OARs are manually delineated by radiation oncologists in the computed tomography (CT) scans. This is usually a time-consuming process due to numerous OARs and often subjects to large inter-operator variability due to the complex shapes of anatomical structures. Therefore, automatic tools that can perform accurate segmentation of OARs can greatly reduce the workloads in treatment planning and guide the radiation dose more optimally distributed.

\begin{figure}[t]
	\centering
	\includegraphics[width=0.49\textwidth,page=1]{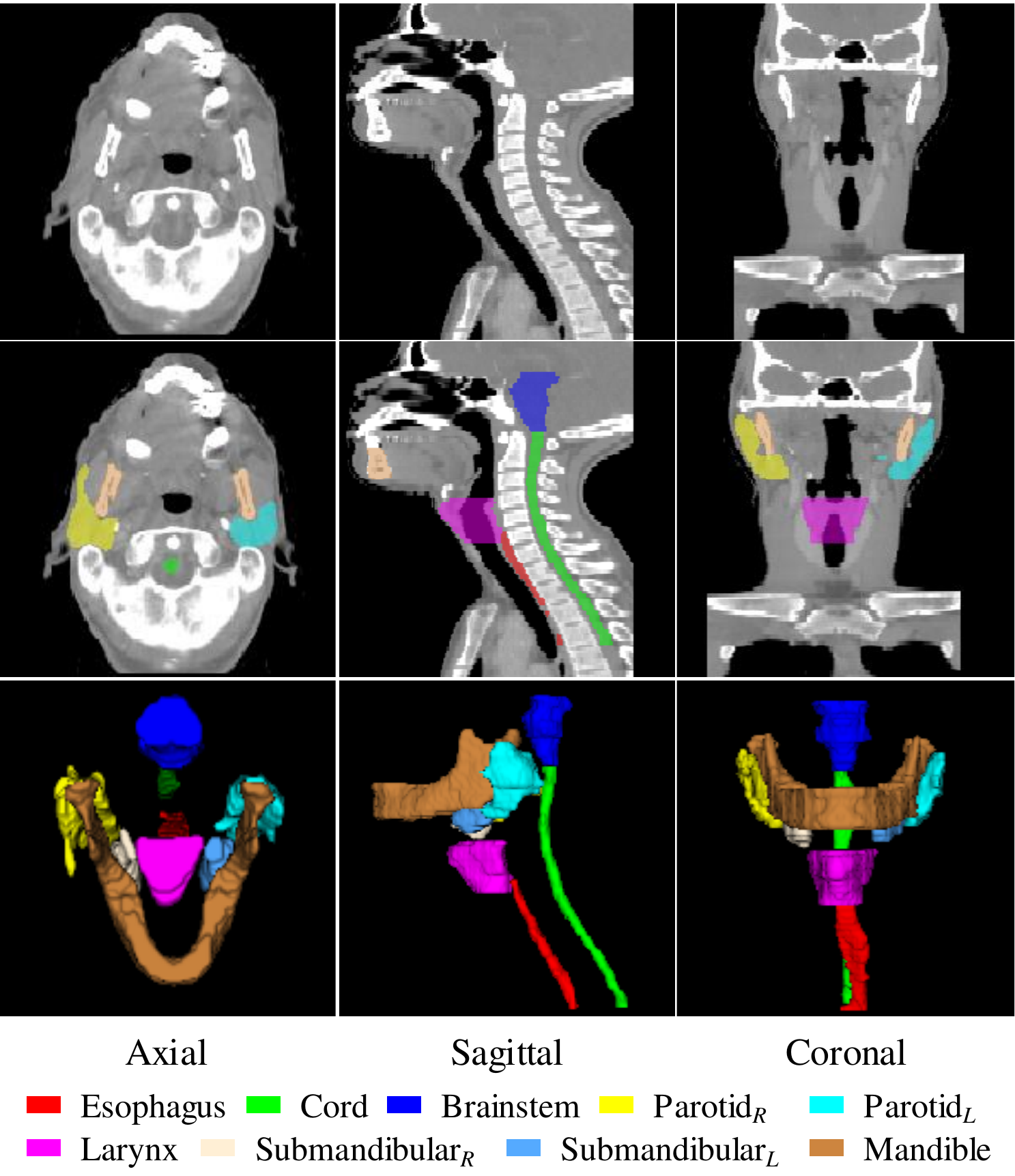}
	\caption{Typical head and neck CT scans and their corresponding segmentations of important OARs. The first, second, and third rows show an original CT slice, the same slice overlaid with segmentation, and a 3D rendering in terms of the axial, sagittal, and coronal views, respectively.}
	\label{fig:Hnn_Example}
\end{figure}

Unfortunately, there are many challenges that limit the performance of automatic segmentation methods on head and neck CT images. \emph{First}, some of the head and neck OARs have low contrast in CT images, which leads to unclear boundaries of these organs and makes them difficult to segment accurately. \emph{Second}, the shape and appearance variability of OARs is high, both between subjects (inter-subject variability) and within subjects (intra-subject variability), which makes it hard to build robust models. \emph{Finally}, the class imbalance problem is serious in head and neck organ segmentation. Especially when the processed OARs are numerous, it is hard to train a universal model to accurately segment all OARs at the same time. 

There are many methods proposed to achieve the accurate segmentation of OARs in head and neck CT images. They can be roughly divided into three categories: multi-atlas-based methods, deformable-model-based methods, and learning-based methods~\cite{chan2019convolutional}. In general, multi-atlas-based methods highly rely on the built templates, which results in these approaches failing in dealing with large anatomical variations~\cite{haq2019dynamic}. For deformable-model-based methods, the main drawback is that they are sensitive to the initialization values so the results are easily biased by noise~\cite{wang2017hierarchical}. In contrast, learning-based methods can more efficiently and effectively learn the mapping from the extracted features to corresponding classes. Instead of manually designing the features in conventional learning-based methods, deep learning-based methods combining feature engineering and classifier into a unified framework for a joint optimization have achieved excellent performance in many medical image segmentation tasks, such as pelvic segmentation~\cite{wang2020iterative}, breast segmentation~\cite{pramanik2019segmentation}, and vessel segmentation~\cite{shi2019intracranial}. But these existing methods usually deal with binary segmentation or limited organ segmentation, if directly applying them to the challenging segmentation of numerous head and neck OARs, the model may be difficult to train and achieve unsatisfactory performance. 

To enhance the capability of the model, many studies are devoted to introducing the organ shape information as an alternative representation of the organ to facilitate the segmentation, such as~\cite{chen2016dcan,shen2017boundary,wang2019ct,wang2020ct}. Most of these methods take the shape regression as an auxiliary task of organ segmentation, implemented by sharing the low-level layers or high-level layers. This multi-task architecture can help improve the segmentation performance, but the gap between different tasks with separate branches make the improvement limited.

To address the challenges outlined above, we propose a novel dual shape guided network (DSGnet) for the segmentation of head and neck OARs in CT images, in which we use an organ-specific unilateral inverse-distance map (UIDM) as the alternative discriminative representation of OARs and utilize it to guide the segmentation in two different perspectives: direct shape guidance and across shape guidance. In the direct shape guidance, the predicted label mask is mapped to the distance space using a simple yet effective encoder-decoder architecture and then supervised by the true UIDM. In the across shape guidance, the prediction for UIDM is as an auxiliary task for guiding the shared feature learning of the segmentation task. Under the proposed framework, the discriminative shape representation can better enhance the model capability from different flows and the gap between different tasks can be alleviated. 

In summary, our main contributions are listed as follows:

\begin{itemize}
	\item We propose a novel dual shape guided fully convolutional network to address the challenging segmentation of numerous OARs in head and neck CT images. Different from conventional multi-task architectures with separate branches, our method can utilize the shape information to direct the segmentation in an end-to-end manner.
	\item We design a new mapping function that transforms the representation from the label space to the distance space using a simple encoder-decoder architecture and introduce it into the segmentation network to make the predicted mask capture the overall shape information.
	\item We build a large dataset with 699 CT scans collected from different patients with head and neck cancers and 179 scans are with nine important OARs annotations by experienced oncologists. To our knowledge, this is the largest head and neck contour dataset in the publication. The comprehensive experiments on this dataset demonstrate that our method can achieve better performance than the published state-of-the-art methods for all nine OARs.
\end{itemize}

\section{Related Work}
Before presenting our new methods to achieve the accurate head and neck OARs segmentation on CT images, we will first briefly introduce the related works using the popular multi-atlas-based methods and also recent learning-based methods for medical image segmentation, respectively.

\subsection{Multi-Atlas-Based Segmentation}
The multi-atlas-based method is the breakthrough for the segmentation of medical images~\cite{wang2012multi,iglesias2015multi,sun2019high}. In general, multiple atlases with labeled ground-truth are first registered with a given image, and then the warped labels from multiple atlases are used as the segmentation results of the given image after the designed fusion strategy. The key techniques for accurate segmentation are to establish reliable registration methods and label fusion criteria.

For example, Jia et al.~\cite{jia2012iterative} constructed a combinative registration tree using both original and simulated images generated using a learned statistical model to improve the registration accuracy. Bai et al.~\cite{bai2013probabilistic} proposed a registration refinement method that incorporating label information into the registration measure to reduce the registration errors for the segmentation in cardiac MR images. \cite{ou2012multi} and \cite{doshi2016muse} selected the optimal atlases from all candidate atlases based on ranking scores to better capture the target anatomy. To improve the correspondence between the atlases and the target images, Asman et al.~\cite{asman2013non} designed a non-local statistical correspondence between an atlas voxel and the given target voxel to resolve imperfect alignment. While the conventional multi-atlas-based methods with weak shape information, Alv{\'e}n et al.~\cite{alven2019shape} utilized landmarks to represent the shape information of the atlases and the target image, and obtained a landmark position map instead of a majority voted probability map to achieve the target image segmentation. For head and neck organ segmentation, multi-atlas-based methods are also widely used. Han et al.~\cite{han2008atlas} developed three hierarchical registration steps with increased degrees of freedom (linear, non-linear, and dense deformable) to gradually improve the registration accuracy. But the multiple registrations pose a serious computational burden. Degen et al.~\cite{degen2016multi} proposed a multi-model local atlas fusion method based on the local similarities and normalized cross correlations to generate the label masks of the synthesized pseudo-CT from MR images. To make the atlas selection and label fusion more adaptive, Haq et al.~\cite{haq2019dynamic} used two weight-attention strategies to dynamically select and weight the atlases for label fusion.

Although multi-atlas-based methods have achieved good segmentation performance in some tasks, their performance highly relies on the existing atlases and the used registration methods, which makes them unable to account for large appearance variations~\cite{leclerc2019deep,wang2020iterative,liu2020ms}. Moreover, the adopted registration process is usually with a high cost in terms of both time and computational resources, which makes them inefficient in real applications. Based on the feedback from oncologists, the performance achieved by the multi-atlas-based method in Raystation (a treatment planning system for radiation therapy of cancer) is quite poor to segment head and neck images.
  
\subsection{Learning-Based Segmentation}
Different from multi-atlas-based methods, learning-based methods give a single label for each voxel based on the learned mapping from the extracted feature representation and no other prior information is required~\cite{kohlberger2011automatic}. Therefore, learning-based methods are more flexible and universal in real applications. There are usually two key technologies to build the learning-based methods - feature engineering and classification model. The feature engineering extracts different features to represent the image/region/voxel and the classification model learns a mapping from the extracted features to the corresponding labels~\cite{cong2015deep}. Based on the feature engineering, the learning-based methods can be divided into two categories: conventional learning-based methods and deep learning-based methods.

The convolutional learning-based methods are based on hand-crafted features. For example, Akselrod-Ballin et al.~\cite{akselrod2006atlas} combined a set of regional features (intensity, texture, and shape features) and the support vector machine (SVM) for the segmentation of brain structures. Mahapatra et al.~\cite{mahapatra2017semi} used low level features to build a self consistency score as the penalty cost in the graph cuts to derive the consensus segmentation. To improve the discrimination of the adopted features, some feature selection methods are designed~\cite{shi2016learning,kharrat2019feature}. Gy{\H{o}}rfi et al.~\cite{gyHorfi2019feature} iteratively selected the features based on their used frequency and contributions to correct decisions. After feature selection, the segmentation process time would be greatly improved. There are two main disadvantages of the conventional learning-based methods: 1) the hand-crafted features require carefully designed under rich experience, which limits their representation capability for challenging tasks; 2) the feature engineering and classifier model are separated, which may lead to sub-optimal.  

Recently, deep learning-based methods have achieved leading performance in medical image segmentation~\cite{wang2018interactive,wang2018deepigeos,tajbakhsh2020embracing,zhang2020generalizing}. Different from the conventional learning-based methods, deep learning-based methods automatically learn hierarchical features from the data itself and jointly optimize the classifier in an end-to-end manner. With various variants of deep learning-based methods, fully convolutional networks (FCNs) are more efficient that can make a dense prediction for the whole image~\cite{long2015fully,fang2019automatic,zhang2020context}. One of the most popular architectures for medical image segmentation is U-net~\cite{ronneberger2015u}, including a contracting path to capture hierarchical context information and an expanding path to reconstruct the detailed information, which is designed for 2D image segmentation. To leverage context from adjacent slices for most 3D volumes in clinical practice, there are some excellent 3D networks proposed based on U-net-liked network architectures, such as 3D U-net~\cite{cciccek20163d} and V-net~\cite{milletari2016v}. For head and neck organ segmentation, Ibragimov et al.~\cite{ibragimov2017segmentation} trained a separate convolutional neural network (CNN) to perform the segmentation of each organ in the head and neck area. While this method only performs one voxel in each forward propagation so it is inefficient for inference. To address the imbalance problem between large and small organs, Gao et al.~\cite{gao2019focusnet} first located the center points of small organs and then used an ROI pooling strategy to focus on the surrounding regions of small organs for segmentation while segmented the large organs as usual. To capture the overall shape information for facilitating the segmentation, Tong et al.~\cite{tong2018fully} constrained the FCN training by incorporating learned latent shape representation using the annotated labels. And Xue et al.~\cite{xue2019shape} first predicted the signed distance map defined based on the distance from voxels to organ contours using a regression task and then mapped the map to organ labels by an approximated Heaviside function. The discriminative shape representation has been utilized in many segmentation tasks~\cite{ni2019elastic,hatamizadeh2019end,wang2018deepigeos}. But these methods usually facilitate the segmentation by learning shared features using a multi-task architecture with separate branches or transforming the learned shape representation to the label representation using some post-processing strategies. The semantic gap between separate tasks limits the performance improvement. Therefore, we aim to design a network that can directly introduce the shape representation in the segmentation task.

\begin{figure*}[t]
	\centering
	\includegraphics[width=0.99\textwidth,page=1]{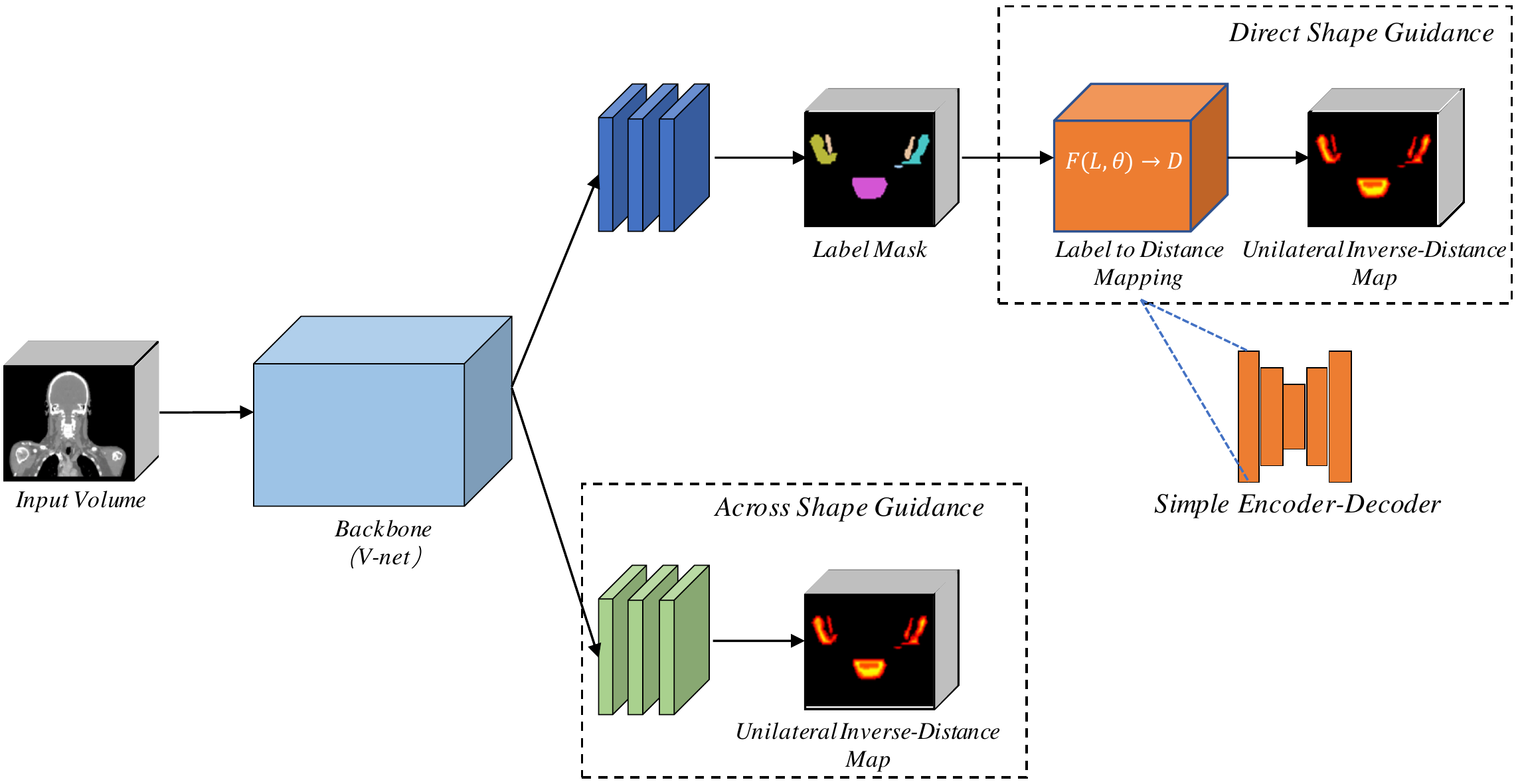}
	\caption{The architecture of our proposed dual shape guided network (DSGnet), in which the direct shape guidance supervises the segmentation prediction by mapping the label space to the distance space using a simple yet effective encoder-decoder module and the across shape guidance enables the model capability by sharing the feature learning using a multi-task architecture. Note that the annotations of only some of nine OARs are shown in this figure.}
	\label{fig:Framework}
\end{figure*}

\section{Method}
In our study, we perform the segmentation for nine important ORAs in head and neck CT images, namely, esophagus, cord, brainstem, left parotid (parotid$_L$), right parotid (parotid$_R$), larynx, left submandibular (submandibular$_L$), right submandibular (submandibular$_R$), and mandible. Therefore, a ten-class (including background) segmentation network is designed. The unclear boundary, large shape and appearance variation, and class imbalance in numerous OARs bring great challenges for the segmentation. To this end, we propose a dual shape guided network (DSGnet) to capture the overall shape from two different perspectives to improve the segmentation performance, the direct shape guidance following the segmentation prediction in an end-to-end manner and the across shape guidance sharing the segmentation feature learning in a multi-task architecture, and the overview of our proposed network is given in Fig.~\ref{fig:Framework}. To represent the organ shape more discriminatingly, a unilateral inverse-distance map (UIDM) is defined based on the distance between the organ voxel and its closest voxel on the target organ contour. To map the representation from the label space to the distance space in the direct shape guidance, a simple yet effective encoder-decoder module is implemented. In the across shape guidance, the regression for the UIDM is used as an auxiliary task to guide the shared feature learning with the segmentation task. In our implementation, V-net~\cite{milletari2016v} is used as the backbone, but any FCN-liked architecture can be fit into our proposed network without additional cost.       

\subsection{Unilateral Inverse-Distance Map}
\begin{figure}[htbp]
	\centering
	\includegraphics[width=0.49\textwidth,page=1]{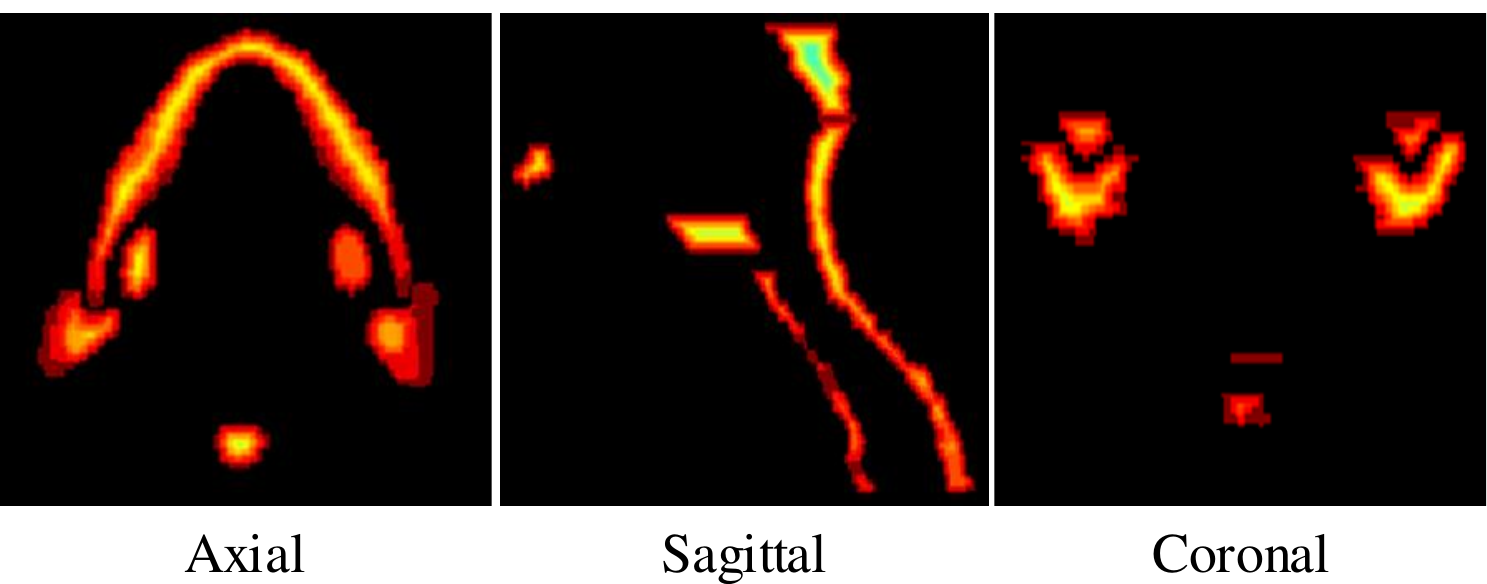}
	\caption{An example of the UIDM from different views.}
	\label{fig:UIDM_Example}
\end{figure}

The main difficulties of head and neck OARs segmentation are their variable shapes across different subjects and unclear boundaries with adjacent tissues. To represent the shape information reliably and robustly, we use an organ-special unilateral inverse-distance map (UIDM) to represent each organ, respectively. Specifically, given a target image and a voxel $x$ belonging to it, the organ-special UIDM is defined:   
\begin{equation}
\psi(x) = \left\{
\begin{aligned}
&0, &x &\in \Omega_{bg}, \\
&exp(-\frac{\min\limits_{y \in S_l}||x-y||_2}{\delta^2}), &x &\in \Omega_l,  
\end{aligned}
\right.
\label{eq:uidm}
\end{equation}     
where $\Omega_{bg}$ denotes the background region, $\Omega_l$ denotes the region belonging to the organ $l$ and $S_l$ is its corresponding contour surface, and $\delta$ is a pre-defined parameter to control the value variation. Based on Eq.~\ref{eq:uidm}, if a voxel on the contour surface of one organ, the corresponding value will be 1, and if a voxel inside one organ, the corresponding value falls in $(0,1)$ which is based on its distance to its closest voxel on the organ contour surface. But if a voxel belongs to the background, its value will be 0. An example of the UIDM is shown in Fig.~\ref{fig:UIDM_Example}. Note that in this definition, the UIDM of each organ is without overlapping with adjacent organs. Moreover, the larger difference on both sides of the contour also makes the representation more discriminative.     

\subsection{Dual Shape Guided Network for Segmentation}
Different from conventional multi-task architectures using separate branches to take the shape representation as a beneficial reference for segmentation, our DSGnet introduces two flows to make the shape representation guide the segmentation: direct shape guidance following the segmentation predictions and across shape guidance sharing the learned features. Under the proposed network, the segmentation model can be enabled to capture the overall shape of the target organ in a more intuitive end-to-end fashion. In the implementation, we adopt 3D V-net~\cite{milletari2016v} as the backbone which can be seen as an extension of the popular U-net~\cite{ronneberger2015u} by incorporating residual connection, 3D convolution, and Dice loss, which has shown outstanding and stable performance in many medical image segmentation tasks.

\subsubsection{Direct Shape Guidance}
\begin{figure}[htbp]
	\centering
	\includegraphics[width=0.49\textwidth,page=1]{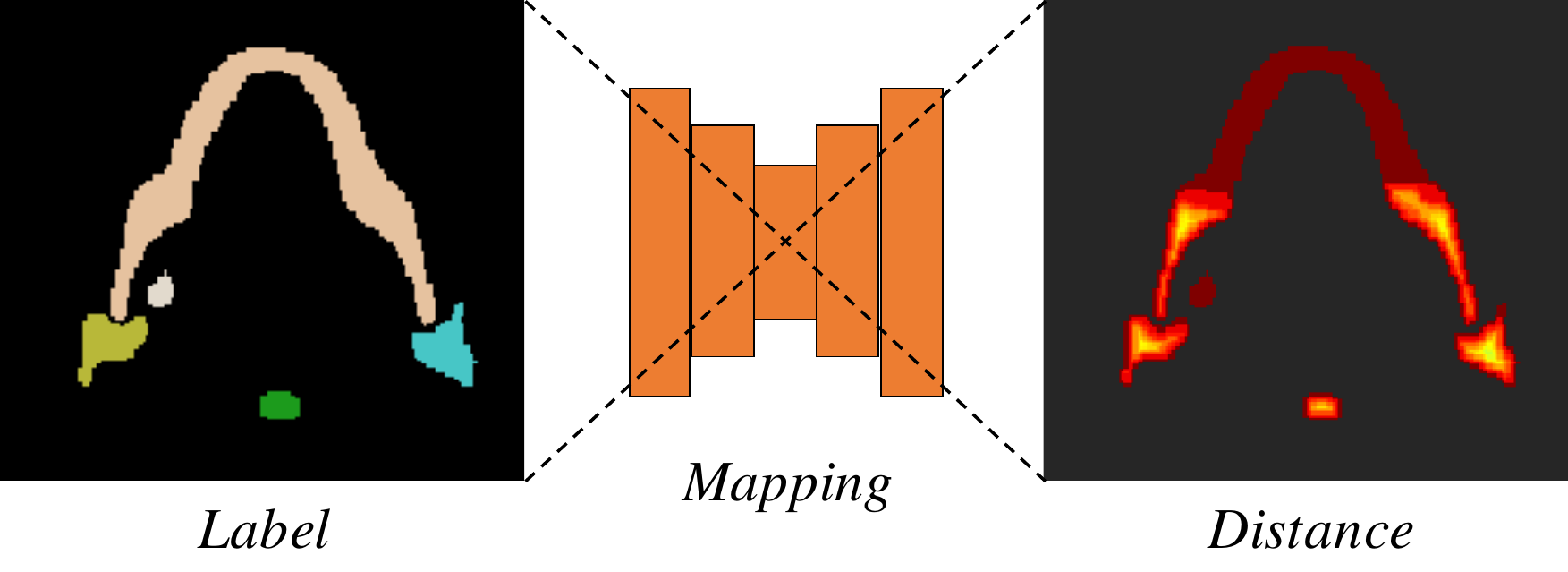}
	\caption{Mapping from the label space to the distance space.}
	\label{fig:Mapping}
\end{figure}

Based on the definition of Eq.~\ref{eq:uidm}, it is not easy to directly extract the predicted UIDM from the predicted label mask and then apply appropriate constraints to capture the overall organ shape while no computable gradient is available for back-propagation. Therefore, in our direct shape guidance, we map the label space for defining the segmentation mask to the distance space for defining the UIDM by a learnable encoder-decoder architecture. That is to say, we assume that Eq.~\ref{eq:uidm} can be fitted by an FCN. Under this assumption, the overall organ shape supervision is implemented in an end-to-end manner, as shown in Fig.~\ref{fig:Mapping}.      

Considering that Eq.~\ref{eq:uidm} is relatively simple, a simple yet effective encoder-decoder network is used to build the mapping. Specially, a simplified version of 3D U-net~\cite{cciccek20163d} is adopted with only two $2 \times 2 \times 2$ pooling layers in the contracting path. And the number of channels in all convolutional layers is 32 except for the last one for prediction. To optimize the mapping function, the mean absolute error (MSE) between the prediction and the true UIDMs is adopted for training. Note that the parameters of this mapping function are first learned to capture the mapping characteristics before training the whole segmentation task.        

After building the mapping from the label space to the distance space, the direct shape guidance branch is optimized with the objective loss function as:

\begin{equation}
\begin{split}
\mathcal{L}_D & = \mathcal{L}_{Dice} + \lambda\mathcal{L}_{MSE}^D \\
&= 1-\frac{2\sum_{i=1}^{n}p_i\hat{p}_i}{\sum_{i=1}^{n}p_i+\sum_{i=1}^{n}\hat{p}_i} + \lambda\frac{\sum_{j=1}^{n}|s_j-\hat{s}_j|}{n},
\end{split} 
\label{eq:direct}
\end{equation}
where $p_i$ and $s_j$ denote the true label and unilateral inverse-distance value of the $i$-/$j$-th voxel, and $\hat{p}_i$ and $\hat{s}_j$ are their corresponding predictions, respectively. The first term constrains the segmentation prediction and the second term constraints the mapped UIDM. And $\lambda$ is the balance parameter for weighting two items. The mapping function is fixed in the segmentation process.
     
\subsubsection{Across Shape Guidance}
The across shape guidance module considers the shape prediction as an auxiliary task to serve as the supplementary guidance for the segmentation and is achieved in a multi-task learning architecture. Different from the direct shape guidance, the across shape guidance plays its role by sharing the learned feature maps with the segmentation task. And only the UIDM is used to optimize this task. Instead of only sharing the low-level features, all convolutional layers are shared except for the last block for final prediction. Similar to the mapping function, MSE is used for optimization as:

\begin{equation}
\mathcal{L}_{MSE}^A = \frac{\sum_{j=1}^{n}|s_j-\hat{s}_j|}{n},
\label{eq:across}
\end{equation} 

\subsubsection{Network Training}        
In order to make full use of the shape representation, we aggregate the supplementary guidance from both direct and across modules. Thus, the final whole objective loss function is:

\begin{equation}
\mathcal{L} = \mathcal{L}_{Dice} + \lambda\mathcal{L}_{MSE}^D + \beta\mathcal{L}_{MSE}^A,
\label{eq:loss}
\end{equation}     

To make the mapping assumption hold in the end-to-end training, the mapping function is first trained using the ground-truth label mask as the input and corresponding UIDM as the supervision. After that, the learned parameters of the mapping function are frozen in the whole network training. And note that the UIDM is just required in the training phase.

\section{Experimental Results}
\subsection{Dataset}
We have collected the head and neck CT scans in collaboration with North Carolina Cancer Hospital to build a large dataset for evaluation. There were a total of 699 images collected from different patients with head and neck cancers, which were scanned during September 2015 and October 2019. While these images were used for radiation therapy planning, some of the head and neck OARs have been carefully delineated and checked by oncologists with expert experience. In our study, we perform multi-class segmentation for nine important OARs (esophagus, cord, brainstem, left parotid, right parotid, larynx, left submandibular, right submandibular, and mandible), so 179 images with all annotations of all nine OARs were used in our experiments. 

The image size ranged from $512 \times 512 \times 90$ to $512 \times 512 \times 383$, the in-plane resolution ranged from $0.518$ $mm$ to $1.367$ $mm$, and the slice-thickness ranged from $1.500$ $mm$ to $3.000$ $mm$. The image intensity ranged from -1024.0 $Hu$ to 3071.0 $Hu$. Before inputting the images for segmentation, spatial normalization and intensity normalization were performed. For spatial normalization, all images and also their corresponding labeled masks were resampled to the same resolution $1 \times 1 \times 2$ $mm^3$. For intensity normalization, we first set all values less than -400 to -400 and values greater than 1200 to 1200, and then performed min-max normalization to normalize all values to [0, 1].

\begin{figure*}[htpb]
	\centering
	\includegraphics[width=0.99\textwidth,page=1]{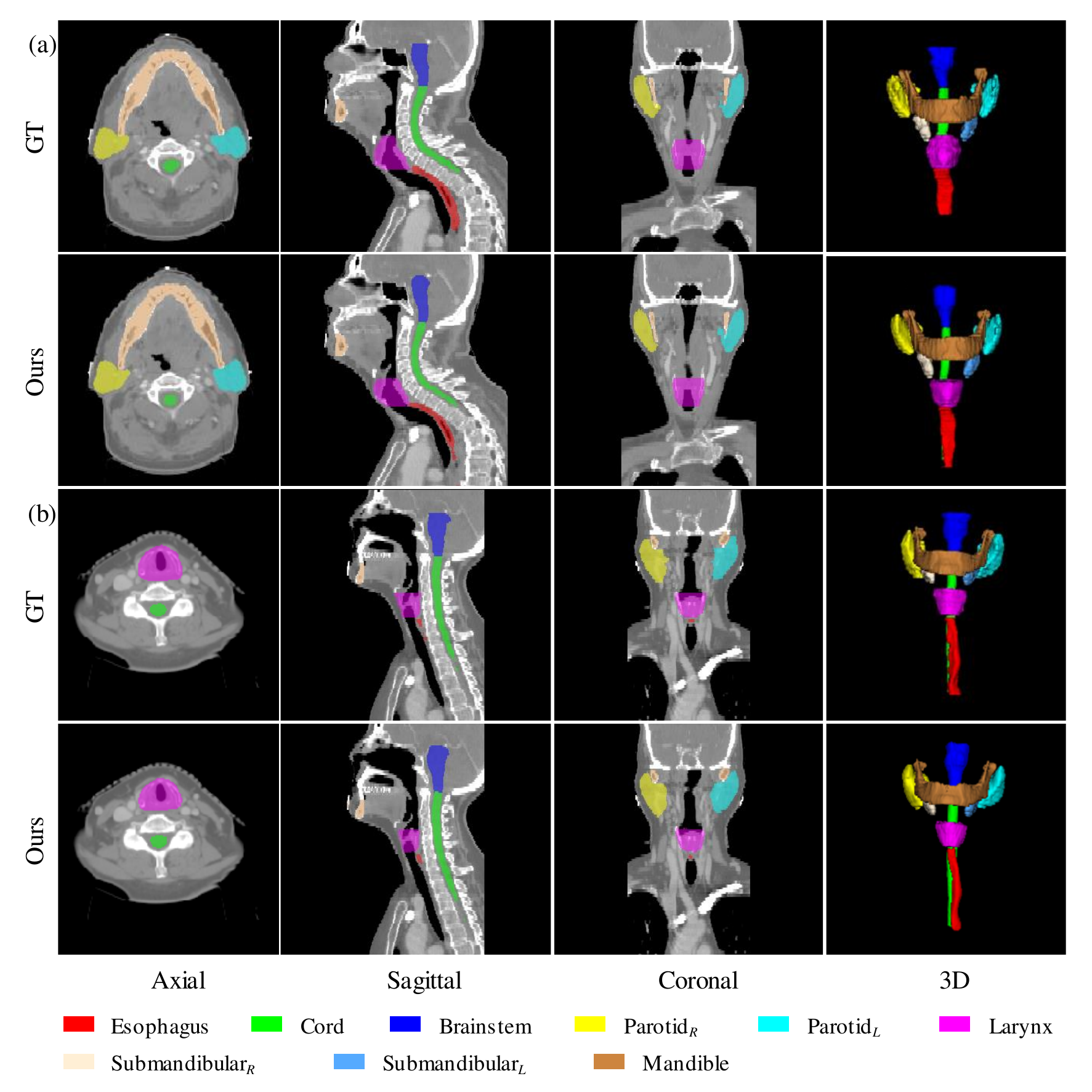}
	\caption{Visual comparison of segmentation results achieved by our method with the ground-truth (GT).}
	\label{fig:Ours_Slice_Example}
\end{figure*}

\subsection{Experimental Settings}    
All implementations in this study are written using Python and all networks are trained using an open-source deep learning library Keras~\cite{chollet2015keras} with TensorFlow~\cite{abadi2016tensorflow} as backend on an Nvidia GTX 1080-Ti graphics card with 11 Gb memory. For making a comprehensive comparison, both 2D and 3D state-of-the-art methods are compared. Specially, the 2D architectures, 2D U-net~\cite{ronneberger2015u}, SegNet~\cite{badrinarayanan2017segnet}, DeepLabv3+~\cite{chen2018encoder}, and DenseNet~\cite{huang2017densely}, and the 3D architectures, 3D U-net~\cite{cciccek20163d}, V-net~\cite{milletari2016v}, and 3D U-net with residual connections (3D RU-net), are compared while these methods have demonstrated excellent performance in many medical image segmentation tasks~\cite{oktay2018attention,hesamian2019deep,tajbakhsh2020embracing}.  
Since the size of our collected dataset is large enough, no data augmentation and fine-tuning are performed for our method and also all compared methods. For all methods, the parameters are initialized using He normal initializer~\cite{he2015delving} and all parameter gradients are clipped to a maximum norm of 1 in the Adam optimization to avoid gradient explosion. The initial learning rate is set 10$^{-3}$ and reduced with a factor 0.1 if validation quantities are not improved after 5 epochs. The patches with size $128 \times 128 \times 3$ and $128 \times 128 \times 48$ are randomly cropped from the original images for 2D methods and 3D methods, respectively. The batch size is 16 for 2D methods and 2 for 3D methods, respectively. 

The popular metrics in medical image segmentation, Dice Similarity Coefficient (DSC) and Average Surface Distance (DSC) are used for evaluation in our experiments~\cite{taha2015metrics}. Three fold cross-validation is performed and the averaged results are reported.                    

\begin{table*}[!tp]
	\renewcommand\arraystretch{1.25}
	\centering
	\caption{Quantitative comparison of segmentation models in terms of mean DSC values with 95\% confidence intervals. The best results are marked as \textcolor[rgb]{1,0,0}{red} while the second best results are marked as \textcolor[rgb]{0,0,1}{blue}. - indicates failed segmentation.}
	\begin{tabular}{c|p{15mm}<{\centering}|p{15mm}<{\centering}|p{15mm}<{\centering}|p{15mm}<{\centering}|p{15mm}<{\centering}|p{15mm}<{\centering}|p{15mm}<{\centering}|p{15mm}<{\centering}}
		\hline \hline
		\diagbox{Organ}{Method} &2D U-net  ~\cite{ronneberger2015u} &SegNet  ~\cite{badrinarayanan2017segnet} &DeepLabv3+ ~\cite{chen2018encoder} &DenseNet  ~\cite{huang2017densely} &3D U-net  ~\cite{cciccek20163d} &V-net~\cite{milletari2016v}  &3D RU-net  &Ours  \\ \hline 
		Esophagus     &0.731$\pm$0.021  &- $\setminus$ -  &0.716$\pm$0.026  &\textcolor[rgb]{0,0,1}{0.738$\pm$0.033}  &0.730$\pm$0.020  &0.725$\pm$0.017  &0.701$\pm$0.024  &\textcolor[rgb]{1,0,0}{0.761$\pm$0.015} \\ \hline 
		Cord          &0.820$\pm$0.014  &0.818$\pm$0.014  &0.808$\pm$0.015  &0.821$\pm$0.013  &\textcolor[rgb]{0,0,1}{0.827$\pm$0.008}  &0.816$\pm$0.010  &0.825$\pm$0.008  &\textcolor[rgb]{1,0,0}{0.831$\pm$0.009} \\ \hline
		Brainstem     &\textcolor[rgb]{1,0,0}{0.858$\pm$0.010}  &\textcolor[rgb]{0,0,1}{0.856$\pm$0.010}  &0.849$\pm$0.011  &0.844$\pm$0.011  &0.717$\pm$0.028  &0.830$\pm$0.008  &0.818$\pm$0.057  &0.853$\pm$0.007 \\ \hline 
		Parotid$_R$   &0.825$\pm$0.014  &0.811$\pm$0.014  &0.803$\pm$0.014  &0.640$\pm$0.035  &0.800$\pm$0.019  &\textcolor[rgb]{0,0,1}{0.829$\pm$0.007}  &0.805$\pm$0.057  &\textcolor[rgb]{1,0,0}{0.841$\pm$0.007} \\ \hline 
		Parotid$_L$   &\textcolor[rgb]{0,0,1}{0.829$\pm$0.015}  &0.804$\pm$0.017  &0.796$\pm$0.017  &0.816$\pm$0.014  &0.820$\pm$0.058  &0.816$\pm$0.009  &0.807$\pm$0.013  &\textcolor[rgb]{1,0,0}{0.838$\pm$0.007} \\ \hline 
		Larynx        &\textcolor[rgb]{0,0,1}{0.850$\pm$0.024}  &0.848$\pm$0.023  &0.840$\pm$0.024  &0.840$\pm$0.022  &0.841$\pm$0.018  &0.834$\pm$0.018  &0.828$\pm$0.021  &\textcolor[rgb]{1,0,0}{0.853$\pm$0.015} \\ \hline 
		Submandibular$_R$    &\textcolor[rgb]{0,0,1}{0.807$\pm$0.020}  &0.772$\pm$0.030  &0.769$\pm$0.022  &0.805$\pm$0.020  &0.794$\pm$0.059  &0.806$\pm$0.015  &0.772$\pm$0.058  &\textcolor[rgb]{1,0,0}{0.818$\pm$0.016} \\ \hline 
		Submandibular$_L$    &0.802$\pm$0.024  &0.781$\pm$0.022  &0.775$\pm$0.020  &0.800$\pm$0.023  &\textcolor[rgb]{0,0,1}{0.806$\pm$0.021}  &0.796$\pm$0.057  &0.786$\pm$0.057  &\textcolor[rgb]{1,0,0}{0.830$\pm$0.013} \\ \hline 
		Mandible      &0.908$\pm$0.010  &0.888$\pm$0.009  &0.871$\pm$0.009  &0.901$\pm$0.011  &0.903$\pm$0.006  &\textcolor[rgb]{0,0,1}{0.909$\pm$0.005}  &0.905$\pm$0.006  &\textcolor[rgb]{1,0,0}{0.918$\pm$0.005} \\ \hline 								  
	\end{tabular}
	\label{tab:Model_Com_DSC}
\end{table*}

\begin{table*}[!tp]
	\renewcommand\arraystretch{1.25}
	\centering
	\caption{Quantitative comparison of segmentation models in terms of mean ASD ($mm$) values with 95\% confidence intervals. The best results are marked as \textcolor[rgb]{1,0,0}{red} while the second best results are marked as \textcolor[rgb]{0,0,1}{blue}. - indicates failed segmentation.}
	\begin{tabular}{c|p{15mm}<{\centering}|p{15mm}<{\centering}|p{15mm}<{\centering}|p{15mm}<{\centering}|p{15mm}<{\centering}|p{15mm}<{\centering}|p{15mm}<{\centering}|p{15mm}<{\centering}}
		\hline \hline
		\diagbox{Organ}{Method} &2D U-net  ~\cite{ronneberger2015u} &SegNet  ~\cite{badrinarayanan2017segnet} &DeepLabv3+ ~\cite{chen2018encoder} &DenseNet  ~\cite{huang2017densely} &3D U-net  ~\cite{cciccek20163d} &V-net~\cite{milletari2016v}  &3D RU-net  &Ours  \\ \hline 
		Esophagus    &1.643$\pm$0.237  &- $\setminus$ -  &\textcolor[rgb]{0,0,1}{1.579$\pm$0.194}  &1.619$\pm$0.266  &4.410$\pm$3.017  &1.878$\pm$0.303  &4.130$\pm$2.102  &\textcolor[rgb]{1,0,0}{1.479$\pm$0.151} \\ \hline 
		Cord         &1.031$\pm$0.144  &1.032$\pm$0.135  &1.062$\pm$0.133  &\textcolor[rgb]{0,0,1}{1.003$\pm$0.121}  &1.006$\pm$0.132  &1.054$\pm$0.144  &1.052$\pm$0.129  &\textcolor[rgb]{1,0,0}{0.942$\pm$0.106} \\ \hline 
		Brainstem    &1.565$\pm$0.347  &\textcolor[rgb]{1,0,0}{1.132$\pm$0.078}  &\textcolor[rgb]{0,0,1}{1.180$\pm$0.082}  &1.234$\pm$0.102  &3.640$\pm$2.053  &1.408$\pm$0.075  &2.044$\pm$0.574  &1.217$\pm$0.061 \\ \hline 
		Parotid$_R$  &1.535$\pm$0.159  &1.620$\pm$0.144  &1.713$\pm$0.156  &2.593$\pm$0.252  &2.521$\pm$1.349  &\textcolor[rgb]{0,0,1}{1.534$\pm$0.080}  &1.753$\pm$0.806  &\textcolor[rgb]{1,0,0}{1.436$\pm$0.090} \\ \hline 
		Parotid$_L$  &\textcolor[rgb]{0,0,1}{1.431$\pm$0.136}  &1.639$\pm$0.143  &1.731$\pm$0.146  &1.521$\pm$0.132  &1.564$\pm$0.163  &1.993$\pm$0.250  &2.692$\pm$0.867  &\textcolor[rgb]{1,0,0}{1.380$\pm$0.064} \\ \hline 
		Larynx       &\textcolor[rgb]{1,0,0}{1.303$\pm$0.201}  &\textcolor[rgb]{0,0,1}{1.358$\pm$0.199}  &1.410$\pm$0.196  &1.515$\pm$0.210  &1.461$\pm$0.154    &1.566$\pm$0.188  &2.505$\pm$0.493  &1.397$\pm$0.159 \\ \hline 
		Submandibular$_R$ &\textcolor[rgb]{0,0,1}{1.185$\pm$0.181}  &1.232$\pm$0.173  &1.253$\pm$0.128  &\textcolor[rgb]{1,0,0}{1.066$\pm$0.119}  &1.546$\pm$0.281  &1.444$\pm$0.674  &1.696$\pm$0.442  &1.343$\pm$0.677 \\ \hline 
		Submandibular$_L$ &1.233$\pm$0.216  &1.194$\pm$0.140  &1.205$\pm$0.110  &\textcolor[rgb]{0,0,1}{1.173$\pm$0.192}  &3.187$\pm$2.618  &1.730$\pm$0.221  &1.785$\pm$0.375  &\textcolor[rgb]{1,0,0}{1.060$\pm$0.566} \\ \hline 
		Mandible     &0.528$\pm$0.062  &0.598$\pm$0.054  &0.686$\pm$0.054  &0.577$\pm$0.070  &\textcolor[rgb]{0,0,1}{0.513$\pm$0.045}  &0.717$\pm$0.106  &0.520$\pm$0.042  &\textcolor[rgb]{1,0,0}{0.438$\pm$0.027} \\ \hline								  
	\end{tabular}
	\label{tab:Model_Com_ASD}
\end{table*}

\subsection{Comparisons with the State-of-the-Art Approaches}
We compare our proposed dual shape guided network (DSGnet) with seven state-of-the-art (SOTA) approaches (including both 2D- and 3D-based architectures) that are widely used in medical image segmentation in terms of the metrics, DSC and ASD ($mm$), respectively. The detailed performance for each head and neck OAR achieved by each method is reported in Table~\ref{tab:Model_Com_DSC} and Table~\ref{tab:Model_Com_ASD}. 

In Table~\ref{tab:Model_Com_DSC}, we can observe that our method can achieve the best DSC performance in eight of nine OARs among all compared methods. And for the brainstem, although 2D U-net~\cite{ronneberger2015u} and SegNet~\cite{badrinarayanan2017segnet} achieve slightly better results than ours, our results are more robust with smaller 95\% confidence intervals. The leading results indicate that our proposed method is more efficient than all compared methods. In all OARs to be segmented, the mandible is relatively easy to segment while its intensity appearance in CT images is more distinguished. Even though all methods can produce relatively satisfactory results, our method still improves the mean DSC value of the best performance (achieved by V-net~\cite{milletari2016v}) in all compared methods by 0.009. Due to numerous OARs in a unified framework, SegNet~\cite{badrinarayanan2017segnet} fails to segment the esophagus from the background and DenseNet~\cite{huang2017densely} only obtains the mean DSC value of 0.640 for the right parotid segmentation. In addition, we also can observe that there are differences in the performance of the left and right parotids (as well as the left and right submandibulars) that demonstrates the large variations of head and neck OARs even for organs of the same type.         
    
In Table~\ref{tab:Model_Com_ASD}, the mean ASD values with 95\% confidence intervals of all methods are reported. Generally, our method achieves the minimum surface distance with the ground-truth in six of nine OARs. For the brainstem, SegNet~\cite{badrinarayanan2017segnet} obtains the best performance but at the cost of the failed segmentation of the esophagus. And for the right submandibular, the best method, DenseNet~\cite{huang2017densely}, obtains uncompetitive results for other OARs, especially for the right parotid. In contrast, our method is very robust to segment all OARs. 

To demonstrate our performance more intuitively, we give two representative examples achieved by our method in Fig.~\ref{fig:Ours_Slice_Example} in terms of both 2D and 3D views. Even though the large shape variations and unclear boundaries make the segmentation very challenging, our method still obtains excellent results consistent with the ground-truth.

\begin{table*}[!tp]
	\renewcommand\arraystretch{1.25}
	\centering
	\caption{Comparisons of segmentation performance achieved by ours segmentation backbone (V-net~\cite{milletari2016v}), with the across shape guidance module (V-net~\cite{milletari2016v} + ASG), with the direct shape guidance module (V-net~\cite{milletari2016v} + DSG), and with both the direct and across shape guidance (Ours) in terms of DSC and ASD ($mm$).}
	\begin{tabular}{c|c|p{16mm}<{\centering}|p{15mm}<{\centering}|c||c|p{15mm}<{\centering}|p{15mm}<{\centering}|c}
		\hline \hline
		&\multicolumn{4}{c||}{DSC}  &\multicolumn{4}{c}{ASD ($mm$)} \\ \hline 
		\diagbox{Organ}{Method}  &V-net~\cite{milletari2016v}  &V-net~\cite{milletari2016v} + ASG &V-net~\cite{milletari2016v} + DSG &Ours &V-net~\cite{milletari2016v}  &V-net~\cite{milletari2016v} + ASG &V-net~\cite{milletari2016v} + DSG &Ours \\ \hline 
		Esophagus    &0.725$\pm$0.017  &0.746$\pm$0.014  &0.740$\pm$0.017  &0.761$\pm$0.015 &1.878$\pm$0.303  &1.633$\pm$0.229  &1.678$\pm$0.232  &1.479$\pm$0.151\\ \hline 
		Cord         &0.816$\pm$0.010  &0.825$\pm$0.008  &0.829$\pm$0.008  &0.831$\pm$0.009 &1.054$\pm$0.144  &0.968$\pm$0.102  &1.009$\pm$0.153  &0.942$\pm$0.106\\ \hline 
		Brainstem    &0.830$\pm$0.008  &0.838$\pm$0.005  &0.844$\pm$0.009  &0.853$\pm$0.007 &1.408$\pm$0.075  &1.382$\pm$0.048  &1.293$\pm$0.070  &1.217$\pm$0.061\\ \hline
		Parotid$_R$  &0.829$\pm$0.007  &0.837$\pm$0.007  &0.836$\pm$0.008  &0.841$\pm$0.007 &1.534$\pm$0.080  &1.473$\pm$0.082  &1.454$\pm$0.083  &1.436$\pm$0.090\\ \hline
		Parotid$_L$  &0.816$\pm$0.009  &0.832$\pm$0.007  &0.832$\pm$0.008  &0.838$\pm$0.007 &1.993$\pm$0.250  &1.685$\pm$0.128  &1.452$\pm$0.074  &1.380$\pm$0.064\\ \hline 
		Larynx       &0.834$\pm$0.018  &0.852$\pm$0.018  &0.846$\pm$0.015  &0.853$\pm$0.015 &1.566$\pm$0.188  &2.576$\pm$1.786  &1.437$\pm$0.128  &1.397$\pm$0.159\\ \hline 
		Submandibular$_R$    &0.806$\pm$0.015  &0.827$\pm$0.013  &0.815$\pm$0.015  &0.818$\pm$0.016 &1.444$\pm$0.674  &1.304$\pm$0.671  &1.366$\pm$0.673  &1.343$\pm$0.677\\ \hline
		Submandibular$_L$    &0.796$\pm$0.057  &0.833$\pm$0.013  &0.823$\pm$0.014  &0.830$\pm$0.013 &1.730$\pm$0.221  &1.343$\pm$0.678  &1.301$\pm$0.667  &1.060$\pm$0.566\\ \hline 
		Mandible     &0.909$\pm$0.005  &0.913$\pm$0.005  &0.916$\pm$0.005  &0.918$\pm$0.005 &0.717$\pm$0.106  &0.494$\pm$0.033  &0.460$\pm$0.039  &0.438$\pm$0.027\\ \hline  
		\hline									  
	\end{tabular}
	\label{tab:Shape_Verifitation_DSCASD}
\end{table*}

\begin{figure*}[htbp]
	\centering
	\includegraphics[width=0.80\textwidth,page=1]{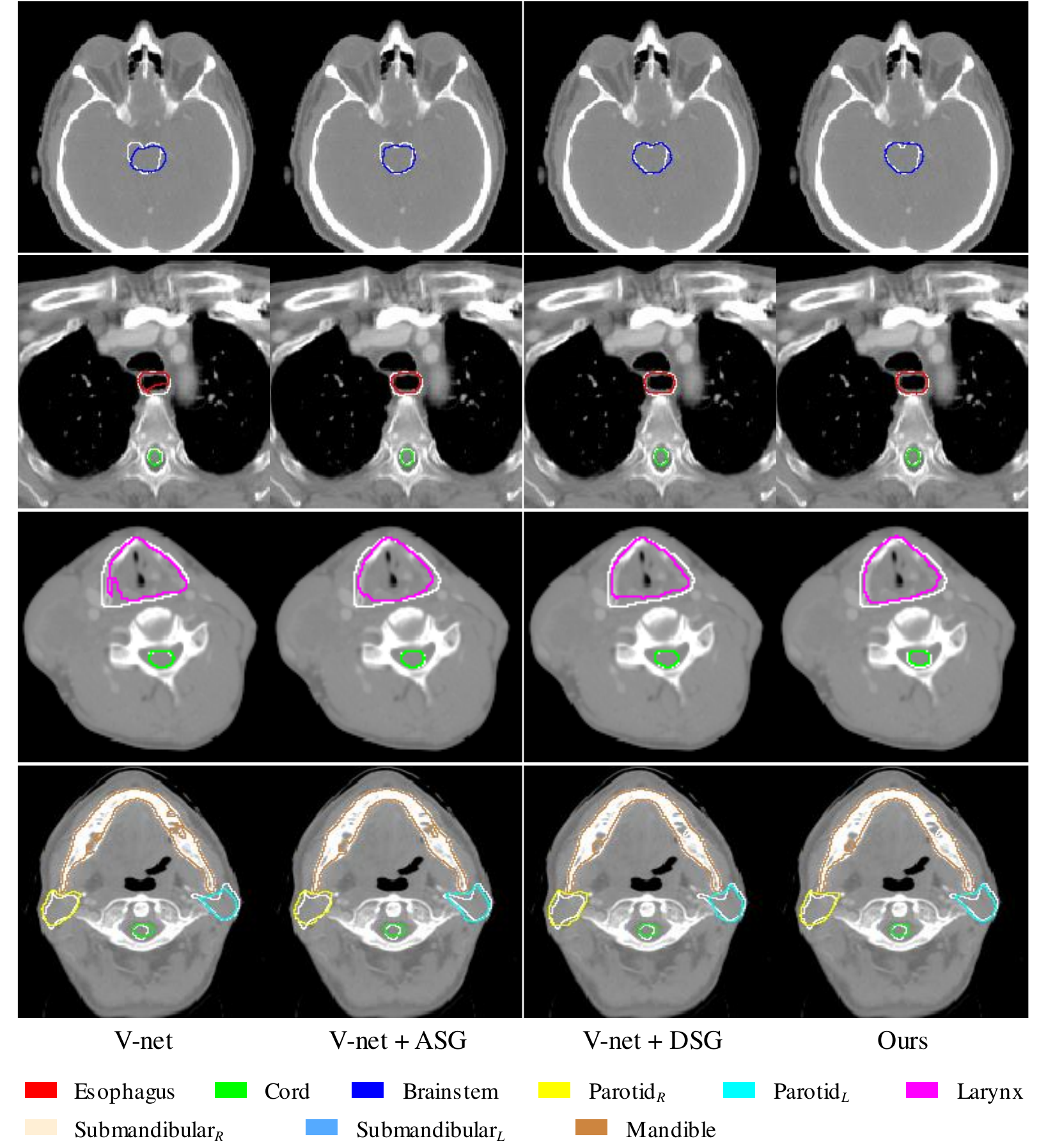}
	\caption{Segmentation results obtained by ours segmentation backbone (V-net~\cite{milletari2016v}), with the across shape guidance module (V-net~\cite{milletari2016v} + ASG), with the direct shape guidance module (V-net~\cite{milletari2016v} + DSG), and with both the direct and across shape guidance (Ours), respectively.}
	\label{fig:SingleTask_VS_Ours_Slice}
\end{figure*}

\subsection{Ablation Study}
\subsubsection{Effectiveness of Shape Guidance}
We introduce a discriminative shape representation to facilitate the segmentation from two different perspectives and implement them with innovative architectures, namely, direct shape guidance following the segmentation prediction and across shape guidance sharing the segmentation feature. To verify the effectiveness of our proposed shape guidance, we implement the following variations to compare with ours: the segmentation backbone with the across shape guidance (V-net~\cite{milletari2016v} + ASG), and with the direct shape guidance (V-net~\cite{milletari2016v} + DSG). Table~\ref{tab:Shape_Verifitation_DSCASD} reports the mean DSC and ASD ($mm$) results with 95\% confidence intervals of these variations and ours. By comparing the performance of V-net~\cite{milletari2016v} and V-net~\cite{milletari2016v} + ASG, we can observe that the shape guidance can improve the segmentation performance of OARs comprehensively even though it only guides the learning of shared features. Instead of guiding the segmentation with a multi-task architecture, V-net~\cite{milletari2016v} + DSG directly constrains the segmentation prediction results to capture the overall shape of OARs and can also improve the performance. The segmentation accuracy can be further improved after introducing both these two shape guidance into the segmentation. These evidences demonstrate that as an alternative representation for OARs, the discriminative UIDM helps the segmentation to capture the overall shape information and makes the results more consistent with the ground-truth.   

\subsubsection{Effectiveness of Mapping Function}
\begin{figure}[t]
	\centering
	\includegraphics[width=0.49\textwidth,page=1]{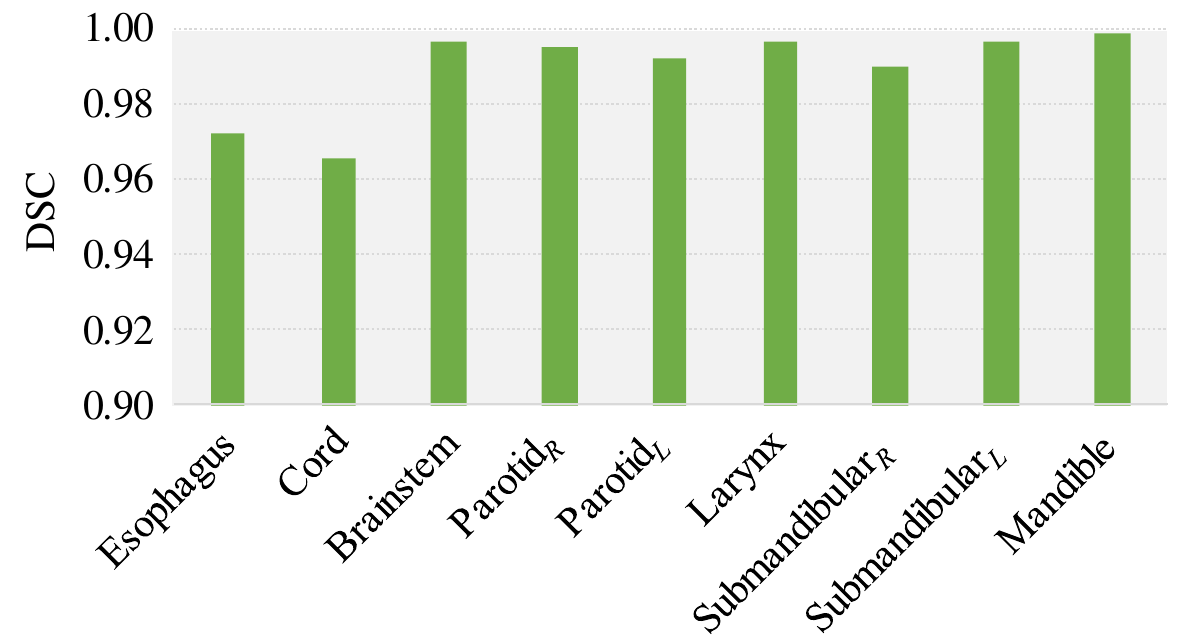}
	\caption{DSC values achieved by the predicted UIDMs using a binarization method on the training set. }
	\label{fig:C2M_Performance}
\end{figure}

In our direct shape guidance module, we realize the mapping transformation Eq.~\ref{eq:uidm} from the label space to the distance space by a learnable simple encoder-decoder architecture. Even though we have verified the effectiveness of our proposed shape guidance module, whether the mapping function can perfectly fit Eq.~\ref{eq:uidm} is yet to be verified. To this end, we use the predicted UIDMs on the training dataset to obtain the predicted label masks using a binarization method with threshold 0. The DSC values of all OARs are shown in Fig.~\ref{fig:C2M_Performance}. We can see that for most OARs, the learned mapping function can perfectly transform the label space to the distance space in terms of DSC values close to 1.

\begin{figure}[t]
	\centering
	\includegraphics[width=0.49\textwidth,page=1]{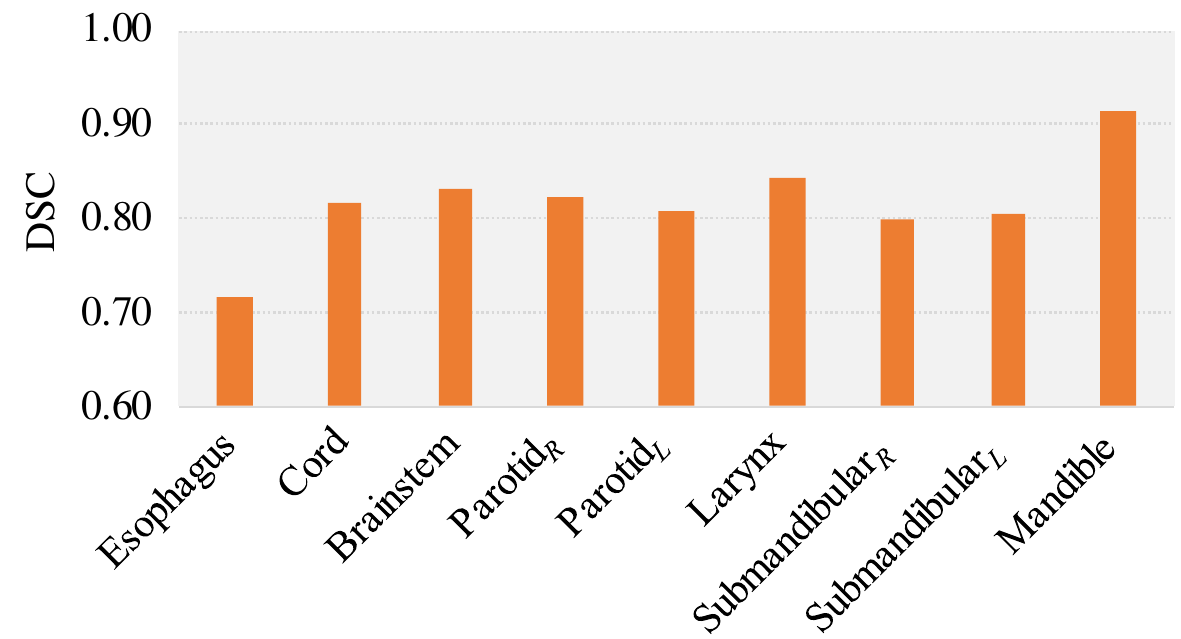}
	\caption{DSC values achieved by first predicting the UIDM and then mapping it to the label mask in an end-to-end manner. }
	\label{fig:Reg_C2M_Performance}
\end{figure}

Moreover, we modify V-net~\cite{milletari2016v} + ASG by exchanging the order of the segmentation and the distance prediction, namely, first learning the UIDM and then mapping it to the label mask by a learnable inverse-mapping function. This implementation is trained in an end-to-end manner and the final segmentation performance is shown in Fig.~\ref{fig:Reg_C2M_Performance}. It is obvious that the segmentation results can achieve almost similar performance by directly using V-net~\cite{milletari2016v}, which verifies that the label space can also be mapped from the distance space using a simple encoder-decoder architecture. That is to say, they are two different representations for OARs with different characterizes that can cooperate to achieve the final segmentation in our method.    

\section{Conclusion}
In this paper, we have proposed a novel dual shape guided fully convolutional network for the segmentation of organs-at-risk in head and neck CT images. We have opened up a whole new manner to utilize the shape representation to facilitate the segmentation. Moreover, the shape guidance mechanism can be easily adapted to other tasks or architectures with no additional overhead. To facilitate subsequent research, we have collected approximately 700 annotated scans, the largest dataset in the publication, from different patients. Comprehensive experimental results show that our results can effectively improve segmentation performance. 
  

%

%
%
%

\ifCLASSOPTIONcaptionsoff
  \newpage
\fi



%

\if false

\fi

\bibliographystyle{IEEEtran}
\bibliography{refs.bib}

\begin{thebibliography}{10}
\providecommand{\url}[1]{#1}
\csname url@samestyle\endcsname
\providecommand{\newblock}{\relax}
\providecommand{\bibinfo}[2]{#2}
\providecommand{\BIBentrySTDinterwordspacing}{\spaceskip=0pt\relax}
\providecommand{\BIBentryALTinterwordstretchfactor}{4}
\providecommand{\BIBentryALTinterwordspacing}{\spaceskip=\fontdimen2\font plus
\BIBentryALTinterwordstretchfactor\fontdimen3\font minus
  \fontdimen4\font\relax}
\providecommand{\BIBforeignlanguage}[2]{{%
\expandafter\ifx\csname l@#1\endcsname\relax
\typeout{** WARNING: IEEEtran.bst: No hyphenation pattern has been}%
\typeout{** loaded for the language `#1'. Using the pattern for}%
\typeout{** the default language instead.}%
\else
\language=\csname l@#1\endcsname
\fi
#2}}
\providecommand{\BIBdecl}{\relax}
\BIBdecl

\bibitem{ACS2020}
A.~C. Society, ``Cancer facts \& figures 2020.''\hskip 1em plus 0.5em minus
  0.4em\relax Atlanta: American Cancer Society, 2020, pp. 1--76.

\bibitem{zorat2004randomized}
P.~L. Zorat, A.~Paccagnella, G.~Cavaniglia, L.~Loreggian, A.~Gava, C.~A. Mione,
  F.~Boldrin, C.~Marchiori, F.~Lunghi, A.~Fede \emph{et~al.}, ``Randomized
  phase iii trial of neoadjuvant chemotherapy in head and neck cancer: 10-year
  follow-up,'' \emph{Journal of the National Cancer Institute}, vol.~96,
  no.~22, pp. 1714--1717, 2004.

\bibitem{chan2019convolutional}
J.~W. Chan, V.~Kearney, S.~Haaf, S.~Wu, M.~Bogdanov, M.~Reddick, N.~Dixit,
  A.~Sudhyadhom, J.~Chen, S.~S. Yom \emph{et~al.}, ``A convolutional neural
  network algorithm for automatic segmentation of head and neck organs at risk
  using deep lifelong learning,'' \emph{Medical physics}, vol.~46, no.~5, pp.
  2204--2213, 2019.

\bibitem{haq2019dynamic}
R.~Haq, S.~L. Berry, J.~O. Deasy, M.~Hunt, and H.~Veeraraghavan, ``Dynamic
  multiatlas selection-based consensus segmentation of head and neck structures
  from ct images,'' \emph{Medical physics}, vol.~46, no.~12, pp. 5612--5622,
  2019.

\bibitem{wang2017hierarchical}
Z.~Wang, L.~Wei, L.~Wang, Y.~Gao, W.~Chen, and D.~Shen, ``Hierarchical vertex
  regression-based segmentation of head and neck ct images for radiotherapy
  planning,'' \emph{IEEE Transactions on Image Processing}, vol.~27, no.~2, pp.
  923--937, 2017.

\bibitem{wang2020iterative}
S.~Wang, Q.~Wang, Y.~Shao, L.~Qu, C.~Lian, J.~Lian, and D.~Shen, ``Iterative
  label denoising network: Segmenting male pelvic organs in ct from 3d bounding
  box annotations,'' \emph{IEEE Transactions on Biomedical Engineering}, 2020.

\bibitem{pramanik2019segmentation}
S.~Pramanik, S.~Ghosh, D.~Bhattacharjee, and M.~Nasipuri, ``Segmentation of
  breast-region in breast thermogram using arc-approximation and
  triangular-space search (bats),'' \emph{IEEE Transactions on Instrumentation
  and Measurement}, 2019.

\bibitem{shi2019intracranial}
F.~Shi, Q.~Yang, X.~Guo, T.~A. Qureshi, Z.~Tian, H.~Miao, D.~Dey, D.~Li, and
  Z.~Fan, ``Intracranial vessel wall segmentation using convolutional neural
  networks,'' \emph{IEEE Transactions on Biomedical Engineering}, vol.~66,
  no.~10, pp. 2840--2847, 2019.

\bibitem{chen2016dcan}
H.~Chen, X.~Qi, L.~Yu, and P.-A. Heng, ``Dcan: deep contour-aware networks for
  accurate gland segmentation,'' in \emph{Proceedings of the IEEE conference on
  Computer Vision and Pattern Recognition}, 2016, pp. 2487--2496.

\bibitem{shen2017boundary}
H.~Shen, R.~Wang, J.~Zhang, and S.~J. McKenna, ``Boundary-aware fully
  convolutional network for brain tumor segmentation,'' in \emph{International
  Conference on Medical Image Computing and Computer-Assisted
  Intervention}.\hskip 1em plus 0.5em minus 0.4em\relax Springer, 2017, pp.
  433--441.

\bibitem{wang2019ct}
S.~Wang, K.~He, D.~Nie, S.~Zhou, Y.~Gao, and D.~Shen, ``Ct male pelvic organ
  segmentation using fully convolutional networks with boundary sensitive
  representation,'' \emph{Medical image analysis}, vol.~54, pp. 168--178, 2019.

\bibitem{wang2020ct}
S.~Wang, D.~Nie, L.~Qu, Y.~Shao, J.~Lian, Q.~Wang, and D.~Shen, ``Ct male
  pelvic organ segmentation via hybrid loss network with incomplete
  annotation,'' \emph{IEEE Transactions on Medical Imaging}, 2020.

\bibitem{wang2012multi}
H.~Wang, J.~W. Suh, S.~R. Das, J.~B. Pluta, C.~Craige, and P.~A. Yushkevich,
  ``Multi-atlas segmentation with joint label fusion,'' \emph{IEEE transactions
  on pattern analysis and machine intelligence}, vol.~35, no.~3, pp. 611--623,
  2012.

\bibitem{iglesias2015multi}
J.~E. Iglesias and M.~R. Sabuncu, ``Multi-atlas segmentation of biomedical
  images: a survey,'' \emph{Medical image analysis}, vol.~24, no.~1, pp.
  205--219, 2015.

\bibitem{sun2019high}
L.~Sun, W.~Shao, M.~Wang, D.~Zhang, and M.~Liu, ``High-order feature learning
  for multi-atlas based label fusion: Application to brain segmentation with
  mri,'' \emph{IEEE Transactions on Image Processing}, 2019.

\bibitem{jia2012iterative}
H.~Jia, P.-T. Yap, and D.~Shen, ``Iterative multi-atlas-based multi-image
  segmentation with tree-based registration,'' \emph{NeuroImage}, vol.~59,
  no.~1, pp. 422--430, 2012.

\bibitem{bai2013probabilistic}
W.~Bai, W.~Shi, D.~P. O'regan, T.~Tong, H.~Wang, S.~Jamil-Copley, N.~S. Peters,
  and D.~Rueckert, ``A probabilistic patch-based label fusion model for
  multi-atlas segmentation with registration refinement: application to cardiac
  mr images,'' \emph{IEEE transactions on medical imaging}, vol.~32, no.~7, pp.
  1302--1315, 2013.

\bibitem{ou2012multi}
Y.~Ou, J.~Doshi, G.~Erus, and C.~Davatzikos, ``Multi-atlas segmentation of the
  prostate: A zooming process with robust registration and atlas selection,''
  \emph{MICCAI Grand Challenge: Prostate MR Image Segmentation}, vol. 2012,
  2012.

\bibitem{doshi2016muse}
J.~Doshi, G.~Erus, Y.~Ou, S.~M. Resnick, R.~C. Gur, R.~E. Gur, T.~D.
  Satterthwaite, S.~Furth, C.~Davatzikos, A.~N. Initiative \emph{et~al.},
  ``Muse: Multi-atlas region segmentation utilizing ensembles of registration
  algorithms and parameters, and locally optimal atlas selection,''
  \emph{Neuroimage}, vol. 127, pp. 186--195, 2016.

\bibitem{asman2013non}
A.~J. Asman and B.~A. Landman, ``Non-local statistical label fusion for
  multi-atlas segmentation,'' \emph{Medical image analysis}, vol.~17, no.~2,
  pp. 194--208, 2013.

\bibitem{alven2019shape}
J.~Alv{\'e}n, F.~Kahl, M.~Landgren, V.~Larsson, J.~Ul{\'e}n, and O.~Enqvist,
  ``Shape-aware label fusion for multi-atlas frameworks,'' \emph{Pattern
  Recognition Letters}, vol. 124, pp. 109--117, 2019.

\bibitem{han2008atlas}
X.~Han, M.~S. Hoogeman, P.~C. Levendag, L.~S. Hibbard, D.~N. Teguh, P.~Voet,
  A.~C. Cowen, and T.~K. Wolf, ``Atlas-based auto-segmentation of head and neck
  ct images,'' in \emph{International Conference on Medical Image Computing and
  Computer-assisted Intervention}.\hskip 1em plus 0.5em minus 0.4em\relax
  Springer, 2008, pp. 434--441.

\bibitem{degen2016multi}
J.~Degen and M.~P. Heinrich, ``Multi-atlas based pseudo-ct synthesis using
  multimodal image registration and local atlas fusion strategies,'' in
  \emph{Proceedings of the IEEE Conference on Computer Vision and Pattern
  Recognition Workshops}, 2016, pp. 160--168.

\bibitem{leclerc2019deep}
S.~Leclerc, E.~Smistad, J.~Pedrosa, A.~{\O}stvik, F.~Cervenansky, F.~Espinosa,
  T.~Espeland, E.~A.~R. Berg, P.-M. Jodoin, T.~Grenier \emph{et~al.}, ``Deep
  learning for segmentation using an open large-scale dataset in 2d
  echocardiography,'' \emph{IEEE transactions on medical imaging}, vol.~38,
  no.~9, pp. 2198--2210, 2019.

\bibitem{liu2020ms}
Q.~Liu, Q.~Dou, L.~Yu, and P.~A. Heng, ``Ms-net: Multi-site network for
  improving prostate segmentation with heterogeneous mri data,'' \emph{IEEE
  Transactions on Medical Imaging}, 2020.

\bibitem{kohlberger2011automatic}
T.~Kohlberger, M.~Sofka, J.~Zhang, N.~Birkbeck, J.~Wetzl, J.~Kaftan,
  J.~Declerck, and S.~K. Zhou, ``Automatic multi-organ segmentation using
  learning-based segmentation and level set optimization,'' in
  \emph{International Conference on Medical Image Computing and
  Computer-Assisted Intervention}.\hskip 1em plus 0.5em minus 0.4em\relax
  Springer, 2011, pp. 338--345.

\bibitem{cong2015deep}
Y.~Cong, S.~Wang, J.~Liu, J.~Cao, Y.~Yang, and J.~Luo, ``Deep sparse feature
  selection for computer aided endoscopy diagnosis,'' \emph{Pattern
  Recognition}, vol.~48, no.~3, pp. 907--917, 2015.

\bibitem{akselrod2006atlas}
A.~Akselrod-Ballin, M.~Galun, M.~J. Gomori, R.~Basri, and A.~Brandt, ``Atlas
  guided identification of brain structures by combining 3d segmentation and
  svm classification,'' in \emph{International Conference on Medical Image
  Computing and Computer-Assisted Intervention}.\hskip 1em plus 0.5em minus
  0.4em\relax Springer, 2006, pp. 209--216.

\bibitem{mahapatra2017semi}
D.~Mahapatra, ``Semi-supervised learning and graph cuts for consensus based
  medical image segmentation,'' \emph{Pattern recognition}, vol.~63, pp.
  700--709, 2017.

\bibitem{shi2016learning}
Y.~Shi, Y.~Gao, S.~Liao, D.~Zhang, Y.~Gao, and D.~Shen, ``A learning-based ct
  prostate segmentation method via joint transductive feature selection and
  regression,'' \emph{Neurocomputing}, vol. 173, pp. 317--331, 2016.

\bibitem{kharrat2019feature}
A.~Kharrat and N.~Mahmoud, ``Feature selection based on hybrid optimization for
  magnetic resonance imaging brain tumor classification and segmentation,''
  \emph{Applied Medical Informatics.}, vol.~41, no.~1, pp. 9--23, 2019.

\bibitem{gyHorfi2019feature}
{\'A}.~Gy{\H{o}}rfi, L.~Kov{\'a}cs, and L.~Szil{\'a}gyi, ``A feature ranking
  and selection algorithm for brain tumor segmentation in multi-spectral
  magnetic resonance image data,'' in \emph{2019 41st Annual International
  Conference of the IEEE Engineering in Medicine and Biology Society
  (EMBC)}.\hskip 1em plus 0.5em minus 0.4em\relax IEEE, 2019, pp. 804--807.

\bibitem{wang2018interactive}
G.~Wang, W.~Li, M.~A. Zuluaga, R.~Pratt, P.~A. Patel, M.~Aertsen, T.~Doel,
  A.~L. David, J.~Deprest, S.~Ourselin \emph{et~al.}, ``Interactive medical
  image segmentation using deep learning with image-specific fine tuning,''
  \emph{IEEE transactions on medical imaging}, vol.~37, no.~7, pp. 1562--1573,
  2018.

\bibitem{wang2018deepigeos}
G.~Wang, M.~A. Zuluaga, W.~Li, R.~Pratt, P.~A. Patel, M.~Aertsen, T.~Doel,
  A.~L. David, J.~Deprest, S.~Ourselin \emph{et~al.}, ``Deepigeos: a deep
  interactive geodesic framework for medical image segmentation,'' \emph{IEEE
  transactions on pattern analysis and machine intelligence}, vol.~41, no.~7,
  pp. 1559--1572, 2018.

\bibitem{tajbakhsh2020embracing}
N.~Tajbakhsh, L.~Jeyaseelan, Q.~Li, J.~N. Chiang, Z.~Wu, and X.~Ding,
  ``Embracing imperfect datasets: A review of deep learning solutions for
  medical image segmentation,'' \emph{Medical Image Analysis}, p. 101693, 2020.

\bibitem{zhang2020generalizing}
L.~Zhang, X.~Wang, D.~Yang, T.~Sanford, S.~Harmon, B.~Turkbey, B.~J. Wood,
  H.~Roth, A.~Myronenko, D.~Xu \emph{et~al.}, ``Generalizing deep learning for
  medical image segmentation to unseen domains via deep stacked
  transformation,'' \emph{IEEE Transactions on Medical Imaging}, 2020.

\bibitem{long2015fully}
J.~Long, E.~Shelhamer, and T.~Darrell, ``Fully convolutional networks for
  semantic segmentation,'' in \emph{Proceedings of the IEEE conference on
  computer vision and pattern recognition}, 2015, pp. 3431--3440.

\bibitem{fang2019automatic}
L.~Fang, L.~Zhang, D.~Nie, X.~Cao, I.~Rekik, S.-W. Lee, H.~He, and D.~Shen,
  ``Automatic brain labeling via multi-atlas guided fully convolutional
  networks,'' \emph{Medical image analysis}, vol.~51, pp. 157--168, 2019.

\bibitem{zhang2020context}
J.~Zhang, M.~Liu, L.~Wang, S.~Chen, P.~Yuan, J.~Li, S.~G.-F. Shen, Z.~Tang,
  K.-C. Chen, J.~J. Xia \emph{et~al.}, ``Context-guided fully convolutional
  networks for joint craniomaxillofacial bone segmentation and landmark
  digitization,'' \emph{Medical Image Analysis}, vol.~60, p. 101621, 2020.

\bibitem{ronneberger2015u}
O.~Ronneberger, P.~Fischer, and T.~Brox, ``U-net: Convolutional networks for
  biomedical image segmentation,'' in \emph{International Conference on Medical
  image computing and computer-assisted intervention}.\hskip 1em plus 0.5em
  minus 0.4em\relax Springer, 2015, pp. 234--241.

\bibitem{cciccek20163d}
{\"O}.~{\c{C}}i{\c{c}}ek, A.~Abdulkadir, S.~S. Lienkamp, T.~Brox, and
  O.~Ronneberger, ``3d u-net: learning dense volumetric segmentation from
  sparse annotation,'' in \emph{International conference on medical image
  computing and computer-assisted intervention}.\hskip 1em plus 0.5em minus
  0.4em\relax Springer, 2016, pp. 424--432.

\bibitem{milletari2016v}
F.~Milletari, N.~Navab, and S.-A. Ahmadi, ``V-net: Fully convolutional neural
  networks for volumetric medical image segmentation,'' in \emph{2016 Fourth
  International Conference on 3D Vision (3DV)}.\hskip 1em plus 0.5em minus
  0.4em\relax IEEE, 2016, pp. 565--571.

\bibitem{ibragimov2017segmentation}
B.~Ibragimov and L.~Xing, ``Segmentation of organs-at-risks in head and neck ct
  images using convolutional neural networks,'' \emph{Medical physics},
  vol.~44, no.~2, pp. 547--557, 2017.

\bibitem{gao2019focusnet}
Y.~Gao, R.~Huang, M.~Chen, Z.~Wang, J.~Deng, Y.~Chen, Y.~Yang, J.~Zhang,
  C.~Tao, and H.~Li, ``Focusnet: Imbalanced large and small organ segmentation
  with an end-to-end deep neural network for head and neck ct images,'' in
  \emph{International Conference on Medical Image Computing and
  Computer-Assisted Intervention}.\hskip 1em plus 0.5em minus 0.4em\relax
  Springer, 2019, pp. 829--838.

\bibitem{tong2018fully}
N.~Tong, S.~Gou, S.~Yang, D.~Ruan, and K.~Sheng, ``Fully automatic multi-organ
  segmentation for head and neck cancer radiotherapy using shape representation
  model constrained fully convolutional neural networks,'' \emph{Medical
  physics}, vol.~45, no.~10, pp. 4558--4567, 2018.

\bibitem{xue2019shape}
Y.~Xue, H.~Tang, Z.~Qiao, G.~Gong, Y.~Yin, Z.~Qian, C.~Huang, W.~Fan, and
  X.~Huang, ``Shape-aware organ segmentation by predicting signed distance
  maps,'' \emph{arXiv preprint arXiv:1912.03849}, 2019.

\bibitem{ni2019elastic}
T.~Ni, L.~Xie, H.~Zheng, E.~K. Fishman, and A.~L. Yuille, ``Elastic boundary
  projection for 3d medical image segmentation,'' in \emph{Proceedings of the
  IEEE Conference on Computer Vision and Pattern Recognition}, 2019, pp.
  2109--2118.

\bibitem{hatamizadeh2019end}
A.~Hatamizadeh, D.~Terzopoulos, and A.~Myronenko, ``End-to-end boundary aware
  networks for medical image segmentation,'' in \emph{International Workshop on
  Machine Learning in Medical Imaging}.\hskip 1em plus 0.5em minus 0.4em\relax
  Springer, 2019, pp. 187--194.

\bibitem{chollet2015keras}
F.~Chollet \emph{et~al.}, ``Keras,'' \url{https://keras.io}, 2015.

\bibitem{abadi2016tensorflow}
M.~Abadi, P.~Barham, J.~Chen, Z.~Chen, A.~Davis, J.~Dean, M.~Devin,
  S.~Ghemawat, G.~Irving, M.~Isard \emph{et~al.}, ``Tensorflow: A system for
  large-scale machine learning,'' in \emph{12th $\{$USENIX$\}$ Symposium on
  Operating Systems Design and Implementation ($\{$OSDI$\}$ 16)}, 2016, pp.
  265--283.

\bibitem{badrinarayanan2017segnet}
V.~Badrinarayanan, A.~Kendall, and R.~Cipolla, ``Segnet: A deep convolutional
  encoder-decoder architecture for image segmentation,'' \emph{IEEE
  transactions on pattern analysis and machine intelligence}, vol.~39, no.~12,
  pp. 2481--2495, 2017.

\bibitem{chen2018encoder}
L.-C. Chen, Y.~Zhu, G.~Papandreou, F.~Schroff, and H.~Adam, ``Encoder-decoder
  with atrous separable convolution for semantic image segmentation,'' in
  \emph{Proceedings of the European conference on computer vision (ECCV)},
  2018, pp. 801--818.

\bibitem{huang2017densely}
G.~Huang, Z.~Liu, L.~Van Der~Maaten, and K.~Q. Weinberger, ``Densely connected
  convolutional networks,'' in \emph{Proceedings of the IEEE conference on
  computer vision and pattern recognition}, 2017, pp. 4700--4708.

\bibitem{oktay2018attention}
O.~Oktay, J.~Schlemper, L.~L. Folgoc, M.~Lee, M.~Heinrich, K.~Misawa, K.~Mori,
  S.~McDonagh, N.~Y. Hammerla, B.~Kainz \emph{et~al.}, ``Attention u-net:
  Learning where to look for the pancreas,'' \emph{arXiv preprint
  arXiv:1804.03999}, 2018.

\bibitem{hesamian2019deep}
M.~H. Hesamian, W.~Jia, X.~He, and P.~Kennedy, ``Deep learning techniques for
  medical image segmentation: Achievements and challenges,'' \emph{Journal of
  digital imaging}, vol.~32, no.~4, pp. 582--596, 2019.

\bibitem{he2015delving}
K.~He, X.~Zhang, S.~Ren, and J.~Sun, ``Delving deep into rectifiers: Surpassing
  human-level performance on imagenet classification,'' in \emph{Proceedings of
  the IEEE international conference on computer vision}, 2015, pp. 1026--1034.

\bibitem{taha2015metrics}
A.~A. Taha and A.~Hanbury, ``Metrics for evaluating 3d medical image
  segmentation: analysis, selection, and tool,'' \emph{BMC medical imaging},
  vol.~15, no.~1, p.~29, 2015.

\end{thebibliography}


\begin{thebibliography}{1}

\bibitem{IEEEhowto:kopka}
H.~Kopka and P.~W. Daly, \emph{A Guide to \LaTeX}, 3rd~ed.\hskip 1em plus
  0.5em minus 0.4em\relax Harlow, England: Addison-Wesley, 1999.

\end{thebibliography}

%

%
%
%




\end{document}